# First-Principles Calculations of the Near-Edge Optical Properties of $\beta$-Ga$_2$O$_3$

Kelsey A. Mengle, Guangsha Shi, Dylan Bayerl, and Emmanouil Kioupakis[a]

*Department of Materials Science and Engineering, University of Michigan, Ann Arbor, Michigan, 48109, USA*

We use first-principles calculations based on many-body perturbation theory to investigate the near-edge electronic and optical properties of β-Ga$_2$O$_3$. The fundamental band gap is indirect, but the minimum direct gap is only 29 meV higher in energy, which explains the strong near-edge absorption. Our calculations verify the anisotropy of the absorption onset and explain the range (4.4-5.0 eV) of experimentally reported band-gap values. Our results for the radiative recombination rate indicate that intrinsic light emission in the deep-UV range is possible in this indirect-gap semiconductor at high excitation. Our work demonstrates the applicability of β-Ga$_2$O$_3$ for deep-UV detection and emission.

The β phase of gallium oxide (β-Ga$_2$O$_3$) is a promising wide-band-gap semiconductor for power electronics, deep-UV optoelectronics, and transparent conductors. Its large band gap results in a high breakdown voltage (8 MV/cm) desirable for power electronics,[1] such as field effect transistors (FETs) and Schottky barrier diodes.[2–4] The large gap also results in visible and ultra-violet (UV) transparency. Solar-blind photodetectors have been successfully fabricated with β-Ga$_2$O$_3$ nanostructures, such as nanowires[5] and nanobelts[6]. These devices demonstrate the potential of β-Ga$_2$O$_3$ in electronic and optoelectronic applications.

Despite numerous experimental and theoretical studies on β-Ga$_2$O$_3$, the nature and value of its fundamental band gap remain controversial. The room-temperature (RT) gap from optical absorption measurements ranges from 4.4 to 5.0 eV.[7] This controversy is partially due to the anisotropy of the crystal structure, which causes the absorption onset to depend on the polarization of the incident light. In addition, the small energy difference between the direct and indirect gaps has led to claims that the fundamental gap is direct.[8,9] Other open questions regard luminescence from this material. Most photoluminescence studies do not show the intrinsic emission across the gap in the deep UV (~265-278 nm) but only emission in the UVA to visible range (~350-600 nm).[10] An exception is the report by Li *et al* (2010) of luminescence at ~265 nm and ~278 nm, which correspond to the experimentally reported absorption edges.[10] Several mechanisms have been proposed to explain the UVA/visible luminescence of β-Ga$_2$O$_3$, including vacancies and other defects and self-trapped hole polarons.[10,11]

---

[1]. Electronic mail: kioup@umich.edu

In this work, we use predictive calculations to investigate the near-edge electronic and optical properties of β-Ga$_2$O$_3$ to resolve open questions about the nature of its band gap. We determine the energies, orbital character, and locations of the band extrema and correlate them with the measured optical properties. We show that β-Ga$_2$O$_3$ is an indirect-gap semiconductor, but the lowest direct gap is only 29 meV larger. To understand the optical properties, we calculate the imaginary part of the dielectric function and the radiative recombination rates at various temperatures and carrier concentrations. Our results explain the different reported optical absorption onsets and demonstrate that intrinsic light emission is possible in this indirect-gap semiconductor.

We performed first-principles calculations using density functional theory and the local-density approximation (LDA) with the Quantum ESPRESSO code.[12] Subsequently, we corrected the band gap with many-body perturbation theory and the G$_0$W$_0$ method using the BerkeleyGW software.[13] We also applied the maximally localized Wannier function method[14] and the Wannier90 code[15] to obtain band structures, dielectric functions, and radiative recombination coefficients[16]. GW calculations used semi-core norm-conserving pseudopotentials,[17] which treat the Ga 3$s$, 3$p$, and 3$d$ electrons as valence states,[18] the generalized plasmon-pole model for the extrapolation of the dielectric function to finite frequency,[19] and the static-remainder approach to converge the sum over unoccupied states.[20] The radiative recombination coefficient was evaluated using Equation (6) in Ref. 16 by interpolating the GW quasiparticle energies and optical matrix elements to fine meshes to sample the small region of the first Brillouin zone occupied by carriers. The band occupations by free carriers were determined using Fermi-Dirac statistics and the frozen-band approximation. The parameters used for the calculations are provided in the Supplementary Material. Calculations were performed using the experimental lattice parameters and atomic positions (Table SI and Figure S1 in the Supplementary Material).[21] The lattice vector magnitudes changed by 0.30-0.68% upon relaxation and resulted in a direct band gap approximately 0.2 eV smaller compared to the unrelaxed structure.

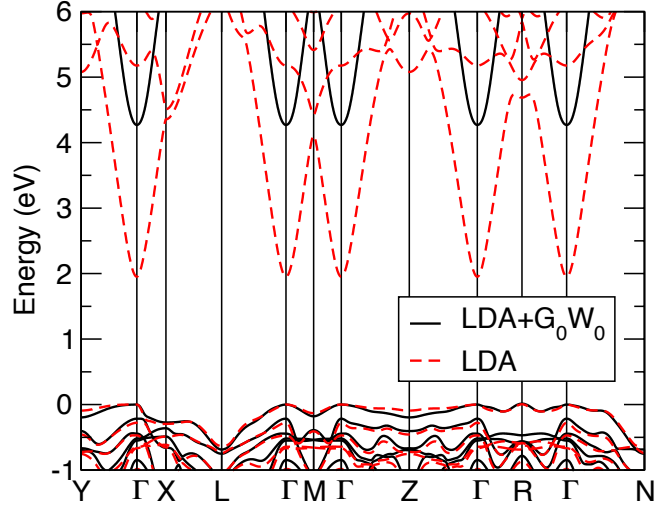

FIG. 1. Near-edge band structure of β-$Ga_2O_3$ calculated with LDA (red dashed) and LDA+$G_0W_0$ (black solid). Both methods show an indirect band gap, the CBM at Γ, and the VBM near R. The LDA (LDA+$G_0W_0$) minimum direct gap is 1.953 eV (4.269 eV), which is 21 meV (29 meV) higher in energy than the indirect gap.

The near-edge band structure of β-$Ga_2O_3$ calculated with LDA and LDA+$G_0W_0$ is shown in Fig. 1 (the band structure over a broader energy range is shown in the Supplementary Material, Fig. S2). The quasiparticle $G_0W_0$ corrections to LDA simultaneously shift up the conduction bands and shift down the Ga 3$d$ bands. It is important to note that the Ga 3$d$ and O 2$p$ orbitals are not hybridized at the LDA level. Such unphysical $pd$ mixing is an artifact of LDA for materials containing cations with shallow $d$ orbitals such as ZnO, InN, and CdO.[22] Since our LDA calculations do not show $pd$ mixing, LDA with semi-core states included in the valence is an accurate starting point for perturbative GW calculations. The topmost valence band of the $G_0W_0$ band structure is composed predominantly of O 2$p$ states with a band width of approximately 7.1 eV. The Ga 3$d$ states are located between -12 and -14 eV below the valence band maximum (VBM), and the O 2$s$ states are located between -17.5 and -19.5 eV below the VBM. The bottom conduction band consists primarily of Ga 4$s$ and O 2$s$ states. The conduction band minimum is located at the Γ-point. Both the LDA and LDA+$G_0W_0$ band structures exhibit the global valence band maximum slightly off the R point along the R—Γ line. Note that the Brillouin zone special points were assigned based on the $TRI_{1b}$ lattice as determined by Setyawan and Curtarolo.[23]

Our LDA+$G_0W_0$ band structure and band gap are in overall agreement with previous theory and experiment (Table I). As expected, LDA severely underestimates the band gap; it predicts an indirect gap of 1.932 eV and a direct gap at Γ that is 21 meV larger. The band gap remains indirect after $G_0W_0$ corrections with a value of 4.240 eV. Our LDA+$G_0W_0$ indirect gap is 29 meV smaller than the direct gap, in comparison to the 30-40 meV range reported in the literature.[7,24–26] This is in contrast to the HSE+GW results of Furthmüller and Bechstedt, which yield a direct gap.[24] The authors of Ref. 24 acknowledge that the discrepancy of their calculated direct gap with the experimentally measured indirect gap could be a result of numerical uncertainties. Our LDA+$G_0W_0$ indirect gap is only 0.2 eV smaller than measured values at RT. We note, however, that the calculated gaps should be compared to experimental data at 0 K, which are approximately 200 meV larger than RT values.[27] Despite the slight gap underestimation, the near-edge band features (indirect gap, symmetry of matrix elements, etc.) are predicted correctly with LDA+$G_0W_0$ and yield accurate near-edge optical properties. Our bulk calculations cannot determine absolute band positions. Reported values for the electron affinity of β-$Ga_2O_3$ range from 2.95 eV to 4.00 eV.[28,29]

TABLE I. The direct and indirect band gaps of β-$Ga_2O_3$ determined by various experimental and theoretical methods.

| Method | $E_{g,\text{indirect}}$ (eV) | $E_{g,\text{direct}}$ (eV) | Reference |
|---|---|---|---|
| LDA | 1.932 | 1.953 | This work |
| LDA+$G_0W_0$ | 4.240 | 4.269 | This work |
| HSE | 4.84 | 4.88 | 25 |
| HSE+GW | 5.046 | 5.038 | 24 |
| Experiment (RT) (polarized transmittance) | 4.43 | 4.48 | 7 |
| Experiment (RT) (ARPES) | 4.85 ± 0.1 | 4.9 ± 0.1 | 26 |

To understand the directionally dependent experimental optical absorption spectra, we need to consider not only the band gap, but also the optical matrix elements for the first several interband transitions. Numerous experiments reported directionally dependent absorption onsets using different polarization angles and crystal orientations.[7,30–34] The different edges arise from several hole bands of different symmetry that reside within 1 eV from the local valence band (VB) maximum at $\Gamma$.[24] Direct optical transitions from the topmost six valence bands to the bottom conduction band at $\Gamma$ occur at 4.27, 4.48, 4.71, 4.79, 4.82, and 5.12 eV, as determined from our LDA+$G_0W_0$ band structure, and correspond to polarization-dependent absorption onsets (Table SII and Fig. 2b). The optical matrix elements show that transitions to the CBM from the third and fifth VB from the top are forbidden. The other four transitions listed above are symmetry allowed. The 4.27 eV transition energy corresponds to polarization along $y$, 4.48 eV to $z$, and 4.79 and 5.12 eV to $x$. The $x$-direction is parallel to the $a$-axis, $y$ is parallel to the $b$-axis, and $z$ is 19.56° off of the $c$-axis. Note that we use different labels for the axes as compared to others in the literature.

The optical matrix elements of the allowed $\Gamma$-$\Gamma$ transitions explain the directionally dependent onset of the calculated imaginary part of the dielectric function (Fig. 2a) since the matrix elements show that each of the bottom four transitions are allowed for only one polarization direction in this material. As seen in Fig. 2, the energies of the absorption onsets match the allowed transition energies for the $\Gamma$-$\Gamma$ transitions. The shoulder around 4.7-5.1 eV along the $x$-direction (also seen experimentally at a smaller intensity[34]) arises from the small but non-zero matrix element of the fourth from the top VB to the CBM, while the larger matrix element of the sixth from the top VB to the CBM dominates at higher energies. The energies on either side of the shoulder also agree with the allowed $\Gamma$-$\Gamma$ transition energies above. Our analysis of the matrix elements and the polarization-dependent absorption edges agrees with the results of Sturm et al.[34] Our results further validate that the measured absorption onset of β-$Ga_2O_3$ can be used to determine the fundamental gap (~4.4-4.5 eV at room temperature[7]) only after considering the anisotropy of this material and the lowest direct absorption onset along the *y* (*b*) crystallographic direction. Optical measurements along different crystal axes and/or non-polarized light result in larger gap values (e.g., 4.7 eV at RT[30]).

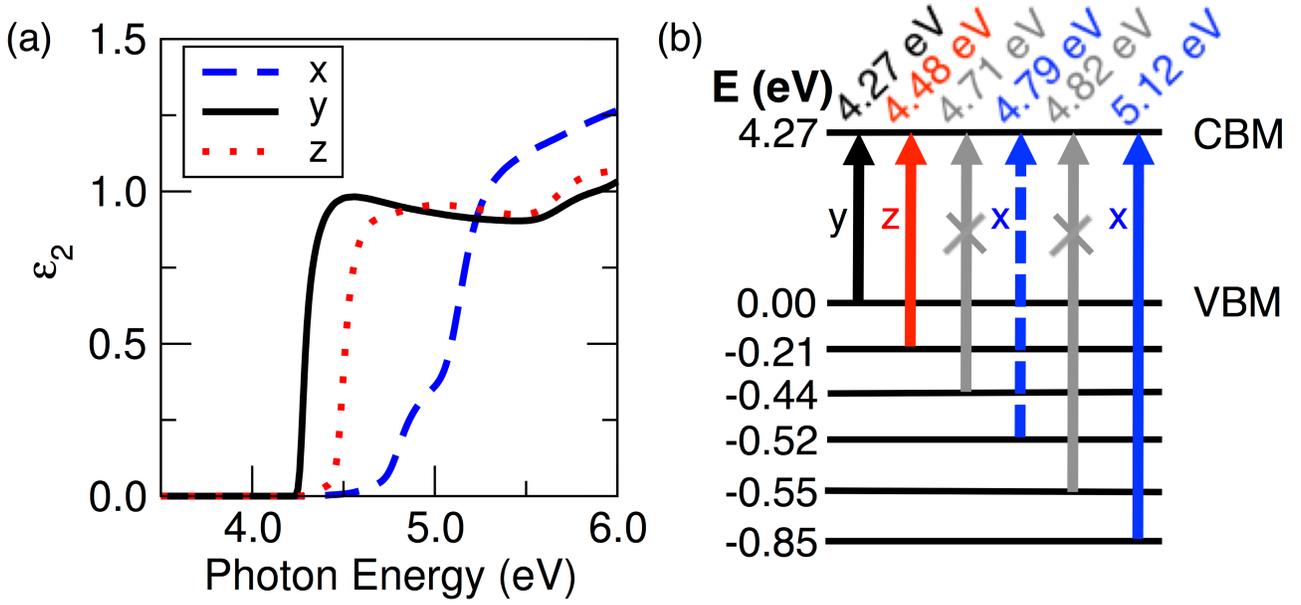

FIG. 2. (a) Imaginary part of the dielectric function along the *x*- (***E*** ∥ ***a***), *y*- (***E*** ∥ ***b***), and *z*- (***E*** ⊥ ***a***, ***b***) directions evaluated with LDA+$G_0W_0$. The absorption onset for light polarized along the *y*-axis (solid black curve) is the minimum LDA+$G_0W_0$ direct gap (4.27 eV). Absorption measurements with light polarized along other crystallographic directions lead to an erroneous overestimation of the band gap. (b) The energies, characters (allowed or forbidden), and polarizations (*x*, *y*, or *z*) of optical transitions at the Γ point explain the directionally dependent onset of the absorption spectra in (a).

Our results also show that, despite having an indirect fundamental gap, β-$Ga_2O_3$ is capable of light emission similar to direct-gap materials. The local VBM at Γ, which corresponds to the lowest-energy direct optical transition, is only 29 meV lower than the global VBM at R. At low carrier concentrations and temperatures, all holes occupy the R valley. As the carrier concentration or temperature increases, holes begin to populate the Γ valley as well. Energy isosurfaces of the top VB are found in the supplementary material (Fig. S3). As the Γ valley fills with holes, light emission becomes possible.

The radiative recombination rate $R_{\text{rad}} = dn/dt = Bn^2$ is proportional to the square of the carrier concentration $n$, where $B$ is the bimolecular radiative recombination coefficient. The magnitude of $B$ was calculated for a range of carrier concentrations and temperatures (Fig. 3). At 4 K and carrier concentrations up to ~$10^{20}$ cm$^{-3}$, $B$ is essentially zero. From 4 K to 200 K, the $B$ coefficient increases by tens of orders of magnitude to values typical of direct-gap semiconductors.[16] Even at 4 K, the $B$ coefficient becomes non-zero as the carrier concentration increases to $10^{21}$ cm$^{-3}$, as degenerate carriers fill the Γ valley. Over the carrier concentration range $10^{15}$–$3.5\times10^{19}$ cm$^{-3}$, the coefficient is maximum for a temperature of 200 K. From approximately $3.5\times10^{19}$–$4\times10^{20}$ cm$^{-3}$, the maximum occurs at 300 K. The $B$ coefficient decreases with subsequent temperature increase, as in direct-gap materials. Fig. 4 shows the temperature dependence of the $B$ coefficient at a carrier concentration of $10^{15}$ cm$^{-3}$. Our results prove that, despite having an indirect band gap, β-Ga$_2$O$_3$ can emit light like a direct-gap material, albeit at a lower recombination rate.

We note that the deep-UV luminescence at photon energies corresponding to the band gap of β-Ga$_2$O$_3$ (4.4-5.0 eV) has not been observed for bulk or thin-film samples. The reported luminescence in the visible range is attributed to defects,[11] while the near-UV luminescence (3.1 eV) is caused by localization of self-trapped hole polarons.[35] However, deep-UV luminescence has been observed in β-Ga$_2$O$_3$ nanowires.[10] One explanation for the emergence of the deep-UV luminescence in nanowires is the reduced concentration of defects.[10] However, self-trapped hole polarons are expected to be stable even in nanowires and would prevent deep-UV luminescence. An additional factor that explains the deep-UV luminescence is the dissociation of hole polarons at high excitation power. The critical density of holes $n_c$ for the Mott transition of trapped hole polarons to free holes is given by [Ref. 36] $n_c^{1/3} \frac{\varepsilon_0 - \varepsilon_\infty}{\varepsilon_0} R_P = 0.25$, where $\varepsilon_0 = 10.2$ is the static dielectric constant [Ref. 37], $\varepsilon_\infty = 3.75$ is the directionally averaged high-frequency dielectric constant [Ref. 24], and $R_P = 2.1$ Å is the bound-state radius, which we assume to be equal to the O-Ga bond length due to the strongly localized nature of hole polarons [Ref. 35]. Our estimate for the critical density for the Mott transition of self-trapped hole polarons to free holes is therefore $4.3\times10^{20}$ cm$^{-3}$, which may be attained in nanowires due to the small sample volume. Our estimates, therefore, indicate that self-trapped hole polarons can dissociate into free holes at high excitation power and recombine with electrons across the gap to produce intrinsic deep-UV photons.

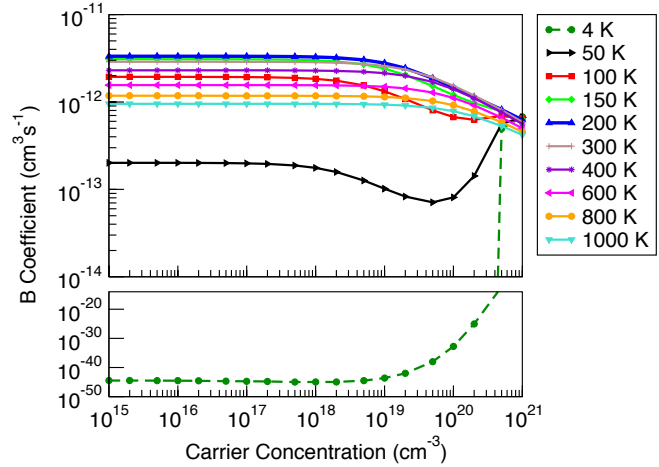

FIG. 3. Bimolecular radiative recombination coefficient *B* of β-Ga$_2$O$_3$ as a function of carrier concentration and temperature. The radiative coefficient drops to essentially zero at 4 K and low carrier concentrations (dashed dark green curve, bottom panel) but increases significantly at higher concentrations or temperatures as holes fill the Γ valley (top panel) to values comparable to direct-gap semiconductors. With increasing temperature between 4 K to 200 K, the low-density ($10^{15}$-$3.5\times10^{19}$ cm$^{-3}$) *B* coefficient increases and subsequently decreases for increasing temperature.

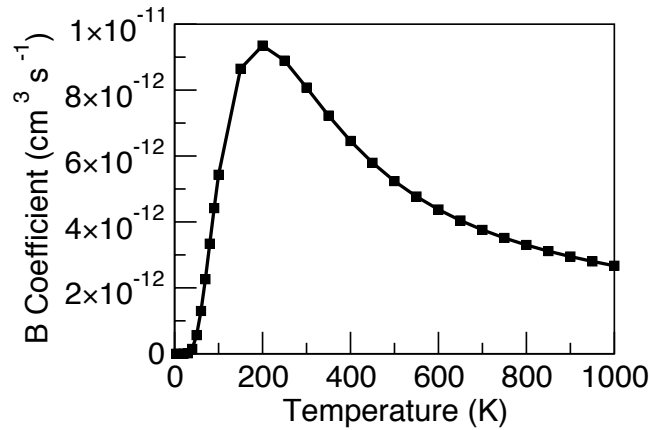

FIG. 4. Low-density ($10^{15}$ cm$^{-3}$) radiative coefficient of β-Ga$_2$O$_3$ between 4-1000 K. The B coefficient is maximum at 200 K before decreasing with increasing temperature as in direct-gap materials.

In summary, we investigated the electronic structure and optical properties of β-Ga$_2$O$_3$ with predictive atomistic calculations. Our results explain the broad range of experimentally reported band-gap values (4.4-5.0 eV at RT) due to the optical anisotropy of the crystal. Linearly polarized light and defined crystal orientations are needed to deduce accurate band-gap values from absorption measurements as in, e.g., Ref. 27. Moreover, our calculations predict that deep-UV luminescence is possible in this indirect-gap material at sufficiently high carrier concentrations and temperatures. Our work elucidates the near-edge optical properties of this important semiconductor for power electronics and deep-UV optoelectronics.

See Supplementary Material for further information and figures on computational details, crystal structure, band structure, and energy isosurfaces of the top valence band.

We acknowledge fruitful discussions with Drs. André Schleife, Joel Varley, Debdeep Jena, and Amit Verma. This work was supported by the Designing Materials to Revolutionize and Engineer our Future (DMREF) program under Award No. 1534221, funded by the National Science Foundation. K.A.M. acknowledges support from the National Science Foundation Graduate Research Fellowship Program through Grant No. DGE 1256260. This research used resources of the National Energy Research Scientific Computing Center, a DOE Office of Science User Facility supported by the Office of Science of the U.S. Department of Energy under Contract No. DE-AC02-05CH11231.

# First-Principles Calculations of the Near-Edge Optical Properties of β-Ga$_2$O$_3$

## Supplementary Material

**Computational Details**

For our density functional theory (DFT) calculations, we used the local-density approximation (LDA) exchange-correlation functional, a plane-wave basis set with a 250 Ry cutoff energy, and a Brillouin-zone sampling grid of 12×6×6. These values converge the total energy to 150 mRy and the electron eigenenergies for states within 6 eV from the band extrema to 1-2 meV. For the GW calculations, the semi-core states were excluded from the parameterization of the generalized plasmon-pole model used to extrapolate the dielectric matrix to finite frequencies,[1] and we used an 8×4×4 Brillouin-zone sampling grid, a screened Coulomb energy cutoff of 45 Ry, and a summation over 3,000 bands. These parameters converge the band gap to within 0.1 eV. We did not consider excitonic effects in the evaluation of the dielectric functions, since excitonic interactions are weakened by phonon screening and dissociate at RT.[2] Radiative recombination coefficients were calculated using Wannier interpolation and a Brillouin-zone sampling grid of 168×108×108.[3]

**Crystal Structure of β-Ga$_2$O$_3$**

The conventional cell of monoclinic β-Ga$_2$O$_3$ belongs to the *C2/m* space group and contains twenty atoms.[4] The gallium ions inhabit two crystallographically distinct positions with half of the ions in an octahedral and half in a tetrahedral geometry. Three different oxygen ion environments exist in the crystal structure, forming a distorted cubic close-packed arrangement. For our calculations we used the primitive unit cell containing four Ga and six O atoms (Fig. S1). This specific primitive cell has a triclinic geometry, where two of the lattice vectors are similar in magnitude. Table SI shows the experimental primitive unit cell lattice vectors and angles used in our calculations. We used the experimental lattice parameters and atomic coordinates without structural relaxation to avoid artifacts by theoretical structural optimization errors on the calculated band structure.

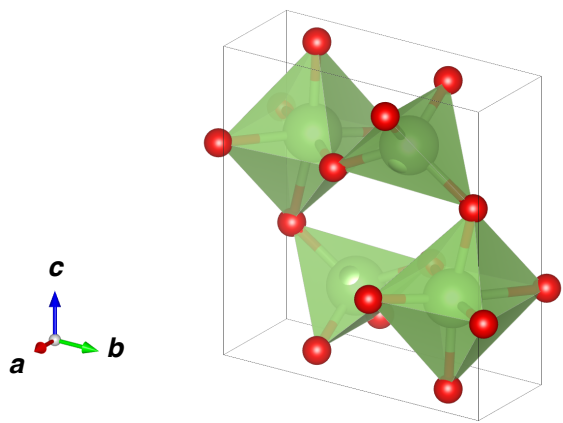

FIG. S1. Primitive unit cell of β-Ga$_2$O$_3$, showing the tetrahedral and octahedral configurations of the Ga ions (large green spheres) with coordinated O ions (small red spheres) used in the calculations.

TABLE SI. Experimental primitive unit cell lattice parameters and angles of β-Ga$_2$O$_3$ used for the calculations.[4]

| | |
|---|---|
| $a$ (Å) | 3.037 |
| $b$ (Å) | 5.798 |
| $c$ (Å) | 6.293 |
| $\alpha$ | 103.414° |
| $\beta$ | 103.964° |
| $\gamma$ | 90.000° |

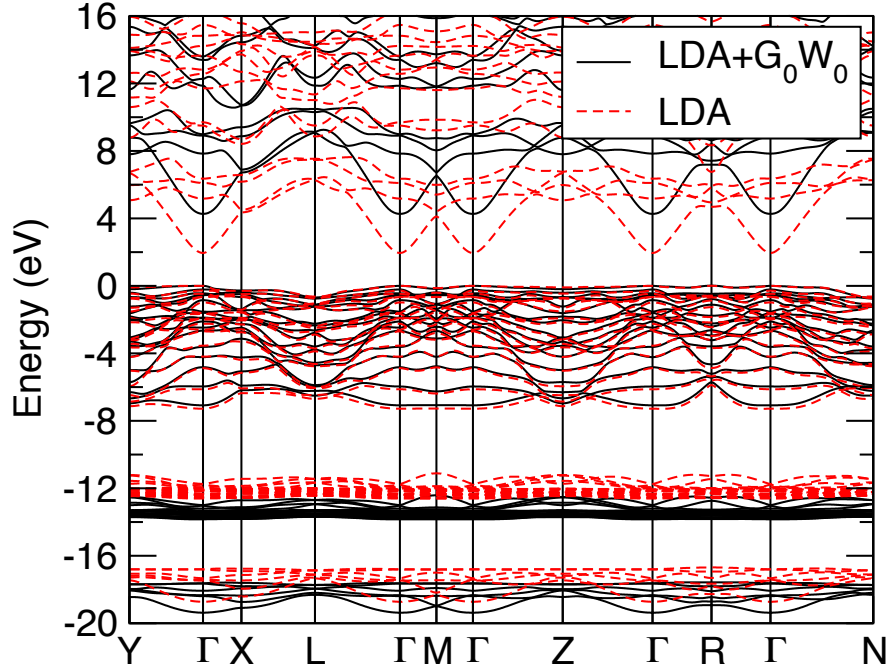

FIG. S2. Band structure of β-Ga$_2$O$_3$ calculated with DFT-LDA (red dashed) and LDA+G$_0$W$_0$ (black solid) methods. Energies are referenced to the top of the valence band.

TABLE SII. Calculated energy differences and optical matrix elements for electronic transitions from the top five valence bands to the conduction band minimum at Γ.

| VB Index | Energy (eV) | Matrix Element (atomic units) |
|---|---|---|
| 1 | 4.27 | 0.73 $\vec{y}$ |
| 2 | 4.48 | 0.73 $\vec{z}$ |
| 3 | 4.71 | 0 |
| 4 | 4.79 | 0.34 $\vec{x}$ |
| 5 | 4.82 | 0 |
| 6 | 5.12 | 0.72 $\vec{x}$ |

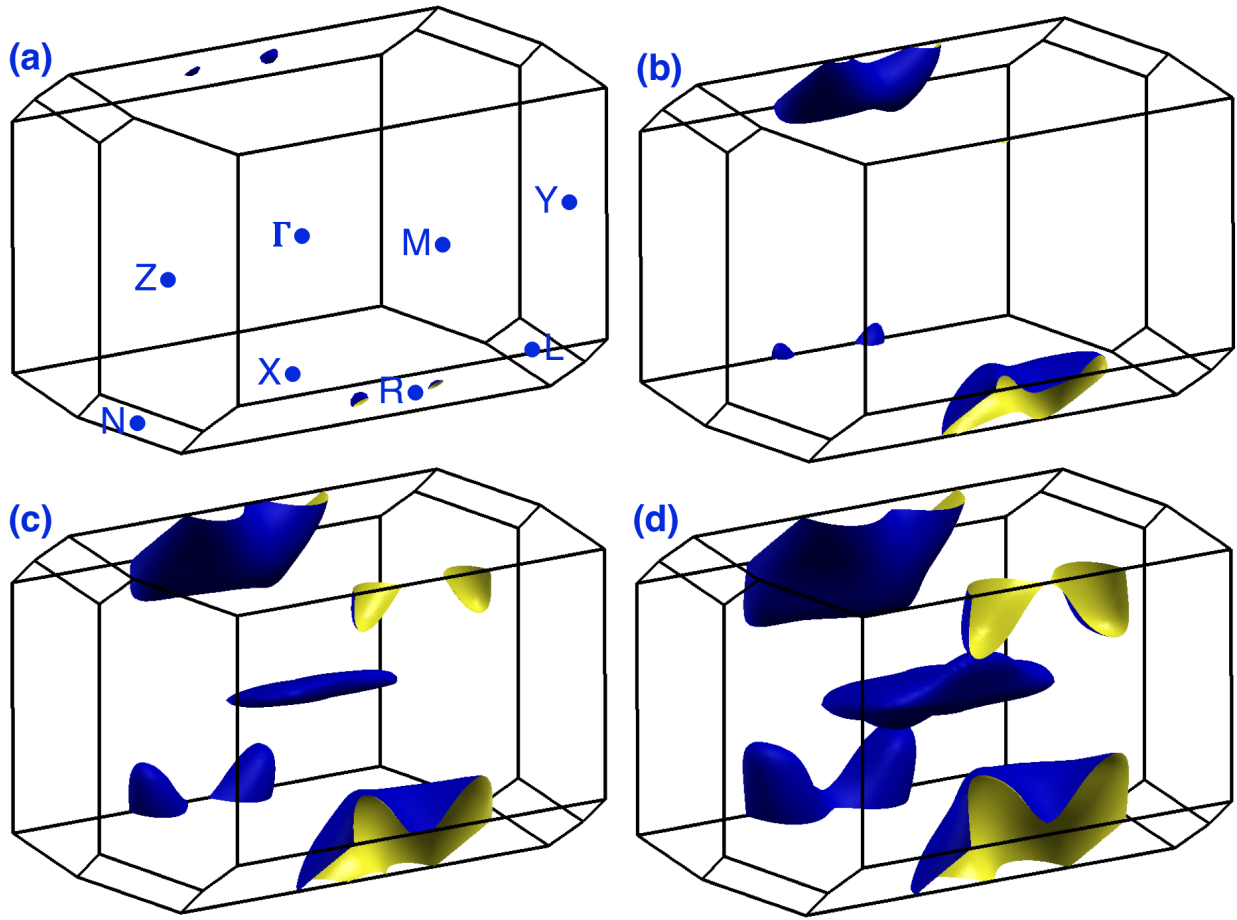

FIG. S3. Top valence band energy isosurfaces of β-Ga$_2$O$_3$ in the Brillouin zone corresponding to the primitive unit cell in Fig. 1 at (a) 0.001, (b) 0.025, (c) 0.050, and (d) 0.075 eV below the VBM. Near the VBM, all holes occupy the R valley. As the energy is lowered to approximately 30 meV below the VBM, holes start to fill the Γ valley. High symmetry points are labeled in (a).